&pdfLaTeX

\documentclass[12pt,preprint]{aastex}

\renewcommand {\deg}   {\mbox{$^\circ$}}

\newcommand   {\kms}   {\mbox{km\,s$^{-1}$}}
\renewcommand {\ga}    {\mbox{\rlap{\hbox{\lower5pt\hbox{$\sim$}}}\hbox{$>$}}}
\renewcommand {\la}    {\mbox{\rlap{\hbox{\lower5pt\hbox{$\sim$}}}\hbox{$<$}}}

\begin{document}



\def\kms {\hbox{km{\hskip0.1em}s$^{-1}$}} 
\def\gos #1 {\left({G\over 10^3 G_0 }\right)^{#1}}
\def\ee #1 {\times 10^{#1}}          
\def\msol{\hbox{$\hbox{M}_\odot$}}
\def\lsol{\hbox{$\hbox{L}_\odot$}}
\def\kms{km s$^{-1}$}
\def\Blos{B$_{\rm los}$}
\def\etal   {{\it et al. }}                     
\def\psec           {$.\negthinspace^{s}$}
\def\pasec          {$.\negthinspace^{\prime\prime}$}
\def\pdeg           {$.\kern-.25em ^{^\circ}$}
\def\degree{\ifmmode{^\circ} \else{$^\circ$}\fi}
\def\ee #1 {\times 10^{#1}}          
\def\ut #1 #2 { \, \textrm{#1}^{#2}} 
\def\u #1 { \, \textrm{#1}}          
\def\nH {n_\mathrm{H}}

\def\ddeg   {\hbox{$.\!\!^\circ$}}              
\def\deg    {$^{\circ}$}                        
\def\le     {$\leq$}                            
\def\sec    {$^{\rm s}$}                        
\def\msol   {\hbox{$M_\odot$}}                  
\def\i      {\hbox{\it I}}                      
\def\v      {\hbox{\it V}}                      
\def\dasec  {\hbox{$.\!\!^{\prime\prime}$}}     
\def\asec   {$^{\prime\prime}$}                 
\def\dasec  {\hbox{$.\!\!^{\prime\prime}$}}     
\def\dsec   {\hbox{$.\!\!^{\rm s}$}}            
\def\min    {$^{\rm m}$}                        
\def\hour   {$^{\rm h}$}                        
\def\amin   {$^{\prime}$}                       
\def\lsol{\, \hbox{$\hbox{L}_\odot$}}
\def\sec    {$^{\rm s}$}                        
\def\etal   {{\it et al. }}                     

\def\xbar   {\hbox{$\overline{\rm x}$}}         

\shorttitle{structure function}
\shortauthors{zadeh}

\title{Imprints of Molecular Clouds in Radio Continuum Images}

\author{F. Yusef-Zadeh}
\affil{Department of Physics and Astronomy
Northwestern University, Evanston, Il. 60208}

\begin{abstract} 
We show radio continuum images of several molecular complexes in the inner Galaxy and report the presence of dark 
features  that coincide with dense molecular clouds. Unlike infrared dark clouds, 
these features which we call ``radio dark 
clouds'' are produced by a deficiency in radio continuum emission from 
molecular clouds that are embedded in a bath of
UV radiation field or synchrotron emitting cosmic ray particles. 
The contribution of the continuum emission along different 
pathlengths results in dark features that trace embedded molecular clouds. 
The  new technique of identifying 
cold clouds  can place constraints on  the depth and the magnetic field of molecular  clouds 
when compared to those of the surrounding hot plasma radiating at radio wavelengths. 
The  study of five 
molecular complexes in the inner Galaxy, Sgr A, Sgr B2, radio Arc, the snake filament 
and G359.75-0.13  demonstrate 
an anti--correlation between the distributions of 
radio continuum and  molecular line and dust emission.
Radio dark clouds  are  identified in GBT maps and  
VLA images taken with uniform sampling of {\it uv} coverage. 
The level at which the continuum flux is suppressed in these sources suggests 
that the depth of the molecular cloud is similar to 
the size of the continuum emission within   a factor of two.
These examples suggest that high resolution, high dynamic range continuum images 
can be  powerful  probes of interacting molecular  clouds with massive stars and supernova remnants 
in regions  where the 
kinematic distance estimates are ambiguous 
as well as in the  nuclei of active galaxies. 
\end{abstract}





\keywords{ISM: clouds ---molecules ---structure Galaxy: center}


\section{Radio Dark Clouds}

The inner few degrees of the Galactic center show a large concentration of molecular, atomic hydrogen and dust clouds 
(Pierce-Price et al. 2000; Lang et al. 2010; Molinari et al. 2011). The molecular gas toward the Galactic center is 
considered to reside in the so-called central molecular zone (CMZ) and consists of a mixture of diffuse and dense 
components (Morris and Serabyn 1996;  Martin et al. 2004; Sawada et al. 2004; Oka et al. 2005; Yusef-Zadeh et al. 2012). Radio continuum 
emission from this region is also extended and is produced by a mixture of thermal and nonthermal processes 
(Nord et al. 2004; Yusef-Zadeh et al. 2004; Law et al. 2008). 
Given the confusing region of the inner Galaxy  due to 
large number of foreground and background sources along the line of sight 
as well as the complex 
motion of the gas in the Galactic center, 
it is difficult to use the kinematic distance method to identify molecular clouds associated 
with HII regions and supernova remnants. We describe a new technique to identify neutral clouds that show a 
deficiency   in the distribution of radio continuum emission. 
These clouds are embedded in a 
bath of radiation or cosmic ray particles produced by thermal or nonthermal sources, respectively. 
The strong  radiation field in the environment of 
cloud  complexes  with high column densities,  such as 
Infrared Dark Clouds (IRDCs), 
allow us to identify their dark counterparts in radio continuum images at cm and mm wavelengths.

The origin of radio dark clouds  (RDCs) is unlike the X-ray shadowing and IRDCs which are caused by strong absorption of 
background light by dense clouds (Egan et al. 1998; Andersen et al. 2010). RDCs are also unlike optically thick 
HII regions seen in absorption against the strong background nonthermal emission at low frequencies. 
This effect is due to a free-free absorption coefficient which increases at low frequencies as             
$\nu^{-2.1}$ (Nord et al. 2004).
The origin of the 
deficiency in radio continuum emission at high frequencies is due to the high column of embedded molecular gas that does 
not allow an external radiation field or cosmic ray particles to penetrate through the cloud. This implies that this subset of 
molecular clouds is interacting with its surrounding hot medium. We first demonstrate the physical situation in which radio 
dark clouds  are produced followed by five examples demonstrating the reality of radio dark clouds.

Since background radio continuum radiation is transparent when passing through neutral clouds, one would expect uniform background 
emission across the face of neutral clouds. However, if a cloud is surrounded by hot synchrotron or thermal 
emitting plasma, the continuum emission is depressed due to the shorter path length of the continuum emission 
integrated along the line
of sight toward the center of the molecular cloud. One possibility involves external 
heating and ionization of neutral  gas by ultraviolet continuum radiation 
that falls off rapidly from the edge to the center 
of the cloud with where the visual extinction (A$_v$) is much larger  than one magnitude. 
Thus, free-free radio emission is substantially reduced in clouds
with high column densities and their imprint can be identified as dark 
features in radio continuum images.  In other words,  atomic or molecular gas clouds or dust 
clouds suppress the continuum emission and create the appearance of a  "hole" 
in their  distribution. 
It is expected that neutral clouds  embedded  in a hot plasma
are  edge brightened outlining the boundary  of the cloud, thus can be distinguished from a
cavity devoid of gas.   The presence of spectral   line and/or continuum dust emission  
can also distinguish RDCs.

We consider a cloud with  a diameter $d$  located at a distance  D from us and is  
embedded within an ionized medium  characterized to have electron  density
$n_e$\,  (cm$^{-3}$) and 
with the  emission measure $E_l=n_e^2\, l$ cm$^{-6}\, pc$ 
along  a path length  $l$.   The 
surface brightness toward the center  and away from  the cloud 
are defined as
S$_{\nu\, ,d}$  and S$_{\nu\, ,l}$, respectively, at a given frequency $\nu$. 
The flux deficiency  $\Delta S_{\nu}$  
is the difference between the flux density 
of the ambient  gas  toward and away from the cloud. 
The ratio of the diameter of the cloud to  the path length $l$ is

\begin{equation}
  \frac{d}{l}  = \frac{\Delta S_{\nu}}{S_{\nu\, ,l}} 
\end{equation}

The differential emission measure $\Delta$EM between the cloud center  and the ionized medium 
can  be estimated by

\begin{equation}
\Delta EM =  \frac{2.15 \ee 2 \Delta S_\nu\,   \nu^{0.1}   T_e^{0.35}}{\theta^2} \, = n_e^2\, d\,\,\,     pc\, cm^{-6}
\end{equation}

where T$_e$ is the electron temperature in 
K, $\theta$ is the beam size in arcsecond and  
$\Delta S_\nu$ is the flux deficiency in mJy.
If the electron density of the ionized medium is measured independently, 
then  the depth  of the molecular cloud $d$ along the line of sight can be estimated.  

Another possibility that 
could produce RDCs
is the deflection of non-thermal particles as they 
diffuse  inside a 
molecular cloud. 
In this case, the path length over which nonthermal particles travel are limited by the magnetic 
field geometry of the cloud which could shield the electrons penetrating into the cloud. 
The ionization losses of nonthermal particles could also 
suppress the emission from high energy 
particles as they  interact  with the gas.  
Thus, the flux of nonthermal emission at 
high frequencies is expected to be 
reduced with respect to the background nonthermal emission. 
If we assume that the magnetic 
field is in equipartition with the particles, the ratio of the magnetic field in the diffuse medium to the 
molecular cloud is

\begin{equation}
\frac{B_l}{B_d} = 
\left(
\frac{S_{\nu\, ,l}\times (l - d)}{S_{\nu\, ,d}\times l}\right)^{\frac{1}{3+\alpha}}
\end{equation}

where $B_l$ and $B_d$ are the magnetic fields in the ambient medium and dense cloud, respectively.  
The spectral index of the emission 
$\alpha$, where S$_{\nu} \propto \nu^{-\alpha}$, is 
assumed to be  constant.

\section{Radio Dark Clouds in the Inner Galaxy}

Multi-wavelength images presented here are based on observations that have already been described elsewhere. 
The data that we have used are taken by Mopra telescope  (Jones et al. 2011), 
Green Bank Telescope (GBT)
of the National Radio Astronomy Observatory\footnote{The
National Radio Astronomy Observatory is a facility of the National Science Foundation, operated
under a cooperative agreement by Associated Universities, Inc.} (NRAO),
 (Law et al. 2008), Very Large Array (VLA) (Yusef-Zadeh et al. 2004), 
Antarctic Submillimeter Telescope and Remote Observatory (AST/RO) (Martin et al. 2004),
IRAC on Spitzer Space Telescope (Arendt et al. 2008), 
SCUBA on James Clark Maxwell Telescope (Pierce-Price et al. 2000), 
Nobeyama Radio Observatory (NRO) (Tsuboi et al. 2011), 
and NICMOS of Hubble Space Telscope (HST) (Yusef-Zadeh et al. 2001). 
We present below multi-wavelength observations of 
five  sources toward the inner Galaxy. 

\noindent {\bf G359.75--0.13:} Figure 1a shows the distribution of velocity integrated HCN (1-0) line emission 
from G359.75-0.13 which runs for $\sim20'$ parallel to the Galactic plane (Jones et al. 2011)). This cloud is 
part of a layer of molecular gas associated with the central molecular zone. This elongated cloud is detected 
at mid-IR and submm images as an IRDC (Molinari et al. 2011; Arendt et al. 2008; Yusef-Zadeh et al. 2009). 
Figure 1b shows the radio continuum counterpart to the molecular cloud at 3.5 cm. A dearth of 
emission coincides with the cloud tracing HCN line emission. Figure 1c shows a composite color image of HCN 
(1-0) emission and 3.5 cm continuum image. Radio continuum emission is present to the north and south of the 
elongated molecular gas layer. To determine the anti-correlation between radio continuum and molecular line 
emission, cross cuts along a line drawn on Fig. 1a are made across the HCN and 3.5 cm radio
 continuum images and are 
presented in Figure 1d. The deficiency in the flux density of radio continuum emission is $\sim$0.1 Jy 
implying that the depth of the ionized gas and molecular gas is similar to each other.  These images 
demonstrate a clear anti-correlation in the distribution of molecular gas that is detected as an IRDC and 
radio dark cloud. The cross cuts combined with images of this cloud at radio and millimeter wavelengths 
suggest that the ionized and molecular gas in G359.75--0.13 are in the same environment, most likely in the 
Galactic center region. This is because diffuse radio continuum emission from the Galactic center is much 
stronger than in the Galactic disk.


\noindent
{\bf G0.13-0.13 and the Radio Arc:}
Adjacent to the nonthermal 
filaments of the Arc near galactic longitude $\sim0.2^{\circ}$ lies the molecular cloud G0.13-0.13
We compare the distribution of radio continuum and
molecular line emission from G0.13-0.13. Figure 2a-b show the distributions of 
a  20cm continuum emission and integrated intensity of HCN (1-0) line, 
respectively. 
The vertical filaments associated with the radio Arc are known to be magnetized 
structures running perpendicular to the Galactic plane. 
A 3.5cm image of the same region mapped by  the GBT shows similar morphology to that of the
20cm continuum image. 
Using HCN line intensity and 3.5 cm continuum images, Figure 2c shows cross cuts,  along a line drawn on Fig. 2b, 
demonstrating the depression in the continuum flux across the length of the cloud. 
The dip at the location of       
G0.13-0.13 corresponds to $\sim700$ mJy at  3.5 cm. 
The molecular cloud G0.13-0.13 is surrounded by vertical 
nonthermal filaments of the Arc to the east (Yusef-Zadeh et al. 1987) 
and to the west of G0.13-0.13 (Reich 2003). The images and cross cuts made from HCN line and radio continuum data  show anti-correlation between molecular line and radio continuum distributions. The 
largest deficiency of radio continuum
emission is located where molecular line emission peaks. 
These  intensity profiles imply  the  interaction of nonthermal radio
filaments and G0.13-0.13 and are  consistent with earlier studies 
measurements (Tsuboi et al. 1997). 
The flux of the continuum emission at 3.5cm is reduced by a factor
2 where the molecular cloud G0.13-0.13 is located. This implies that  the region of nonthermal emission 
surrounding RDC is  twice the depth of the molecular cloud.




\noindent
{\bf G359.16--0.04 and the Snake Filament:} 
One of the most prominent nonthermal filamentary structure in the Galactic center region is the Snake filament 
which extends for more than 20$'$ and runs almost perpendicular to the Galactic plane (e.g., Gray et al. 
1995). Figure 3a shows the northern half of this striking filament terminating at the
 radio continuum source 
 G359.16-0.04  (Uchida et al. 1996). 
This 6cm continuum image  shows two  dark features RDC-1 and RDC-2 to the E. and
W.  edges of the extended continuum source giving the appearance of a ``butterfly".
RDC-2 lies to the north of the well-known 
radio jet 1E1740-2942 (Mirabel et al. 1992). 
Given that the 6cm data is produced by the VLA 
and that incomplete {\it uv}  coverage may be responsible for producing dark features, 
Figure 3b shows a grayscale 3.5cm continuum
emission mapped by the GBT. 
We note that the dark features are also identified in single dish observations, thus establishing
the reality of RDCs in interferometric images.

Contours of integrated emission of CO (4-3) line  are superimposed on the
6cm continuum image, as shown in Figure 3c. 
Given the mismatch between the 2$'$ resolution of molecular line data based on 
AST/RO observations and radio 
continuum data, 
the distribution of molecular gas is similar to that of dark radio clouds. 
The continuum image reveals  detailed morphological structures of a giant molecular cloud on 
arcsecond spatial resolution especially at the boundary of the molecular cloud near 
the continuum source. We also made   cross cuts  at b=$-3'\, 45''$  
across the 6cm and molecular line images, as shown on Figure 3d.
The  cross cuts show two peaks of molecular line emission coinciding with 
 two dips in the 6cm image with  deficient radio flux
of roughly 400 and 300 mJy. These dips   
correspond to RDC-1  and RDC-2, and  
imply  that  the diameters of RDC-1 and RDC-2 are 
roughly 4/5 and 3/5  of the depth of the ionized  medium surrounding the cloud. 


\noindent
{\bf G0.6-0.0 Sgr B2:} 
The  Sgr B2 cloud is a well-studied giant molecular cloud which lies near 
the Galactic center (Reid et al. 2009)  
and is part of a continuous dust ridge that is viewed  in absorption at mid-IR wavelengths (Lis \&
Carlstrom 1994; Molinari et al. 2011). 
Figure 4a shows the  IRDC  associated with Sgr B2 
to the N. and NE  of   the cluster of stars and   the nebula.  Figure  
4b,c show the distribution of 20 and 6cm continuum emission from the same region,  respectively.
The brightness of the 20cm continuum emission is saturated to bring out an elongated dark feature 
near b$=0^\circ$. 
Radio    continuum images
show  RDC-1  which appears to
coincide with the southern edge of  IRDC that surrounds the  cluster of UC HII regions 
in Sgr B2 (De Pree et al. 2000). 
We also note  a decrease in the radio flux at 6 and 20cm between Sgr B2 and
the isolated radio continuum feature G0.73-0.10. However, the image quality in this region 
is poor, thus may show artifacts. 
Cross cut plots along a line drawn on Fig. 4c  across the RDC-1 are made using the  6cm
continuum and 450$\mu$m  images. 
The profiles of emission, as shown in Figure 4d, 
show flux  deficiencies of 5 and 2 mJy   which imply   that 
the depths of the ionized gas for the two dips 
are similar to  and twice larger than 
those  of molecular gas, respectively. 
Because of the  large number of bright compact HII regions in Sgr B2,
the cross cuts also show fluctuations  due to  emission from individual compact sources. 


\noindent
{\bf Sgr A East SNR \& Sgr A West:} 
Given its gaseous  environment 
with a strong radiation field, 
the inner ten parsecs of the Galactic center
is an excellent site 
to search 
for RDCs tracing molecular clouds that interact with the strong radiation field or
with expanding supernova remnants. 
The strong radiation field in the 
Galactic center produces a thick layer of ionized gas at the edge of dark clouds.  
One of the most prominent giant molecular clouds in the Galactic center is the 50 \kms\,  G0.02-0.07. This cloud is physically 
interacting with the expanding shell of the Sgr A East SNR G0.0+0.0 which is thought to lie near the Galactic center (e.g., 
Tsuboi et al. 2011). This complex region is also the site of young massive star formation as  a chain of compact HII regions that 
lie to the east of the Sgr A East remnant. Figures 5a-b show the inner 
4$'$ of the 
Galactic center at 6cm 
hosting the shell-type SNR G0.0-0.0 and four compact HII regions A-D at its   eastern boundary  (Zylka 
et al. 1992). The 1.3mm continuum distribution 
traces dust emission from the 50 \kms molecular cloud. We note several dark 
features  RDC-1 to 4 surrounding the SNR shell of Figure 5a, all of which coincide with dust and molecular features 
(Tsuboi et al. 2011). 
The largest scale dark cloud RDC-1 is 
noted to the NE of the chain of HII regions where we note an oval-shaped structure with an extent of 2$'\times1'$.1. 
Figure 5c shows contours of integrated emission of SiO (2-1) superimposed on the 6cm image, supporting the 
suggestion that RDC-1 coincides  with an oval-shaped molecular gas. Another chain of HII regions
 to the  NE of RDC-1  near $\delta=-28^0\, 58'$ might be associated with a second site of star formation in this cloud. 
To estimate the level at which the continuum emission is suppressed, Figure 
5d shows cross cuts of RDC-3 
made at 1.3mm and 6cm,  supporting  a clear anti-correlation between dust and radio continuum emission. 
The largest 
deficient flux 
at 6cm is 25 mJy implying that the size of the hot plasma is twice the size of the molecular cloud.



On a smaller scale, the inner pc of the Galactic center hosts diffuse ionized gas Sgr A West
orbiting the massive black hole (e.g., Ferriere et al. 2012). Sgr A West consists of three arms 
of ionized  gas and is surrounded by the Circumnuclear molecular ring (CMR). 
Figure 5e,f  show a 3.5cm  and 1.87$\mu$m continuum  
images of the eastern half of 
Sgr A West 
whereas Figures 5g   show  
molecular H2 1-0 S(1) counterparts (Yusef-Zadeh et al. 2001). 
Several dark 
features are noted in the  3.5cm continuum image, especially in the region between the N. and E. arms. 
It turns out RDC-1 and RDC-2 lying  between the two arms 
coincide with an  extinction feature,  
as revealed in Figure 5h where a drop in stellar density is noted. 
Cross cuts across the 3.5cm and H2 images show that the   H2 emission coincides with a dip in the 
continuum peak  at 3.5cm. 
The deficient flux at 3.5cm suggests that the depth of the ionizing gas is similar to that of the 
molecular gas.  From equation (2), the depth of the cloud is estimated to be 0.11 pc if 
$n_e\sim5000$ cm$^{-3}$ and T$_e\sim6000$K.

The association of a "tongue" of neutral gas  (RDC-3),  an extinction cloud 
with  the N.  arm of Sgr A West (Jackson et al. 1993; Yusef-Zadeh et al. 2001), 
as well as 
the association of the E. arm with neutral gas suggest that there is considerable  
molecular gas in the ring and that the arms of Sgr A West 
trace the ionized surface of neutral   clouds.  
If these clouds are massive, 
the presence of molecular gas inside the CMR
 can  have  important consequences  in 
the formation of stars and the dynamics 
of stars near Sgr A*. 


\noindent {\bf Conclusions:} 
We have presented 
five examples of dark features in radio continuum images of molecular complexes toward the inner Galaxy. 
We illustrated  that these dark features 
are anti-correlated with molecular line and dust 
emission, thus implying that cold gas 
associated  with radio continuum features can be detected in 
radio images. 
Given the new generation of radio telescopes with their 
broad band capability, future continuum measurements  could potentially be effective in 
identifying cold gas clouds in  star forming sites  in the local and distant universe. 
Some of the dark radio clouds are surrounded by nonthermal radio emission, 
thus,   radio continuum imaging can identify 
the interaction of  nonthermal particles   from  radio jets or jet-driven outflows  or supernova remnants 
with the cold ISM or the  IGM. Lastly, RDCs can potentially be used as OFF  positions 
in total power technique of observations with  radio telescopes. 
 

Acknowledgments:
I am grateful to  M. Wardle,  D. Roberts, R. Arendt, W. Cotton,  
M. Royster, M. Tsuboi and other colleagues  for discussions and for 
providing  me with their data over the years.    
This work is partially supported by the grant AST-0807400 from the NSF.

\newcommand\refitem{\bibitem[]{}}


\begin{figure}
\center
\includegraphics[scale=0.45,angle=0]{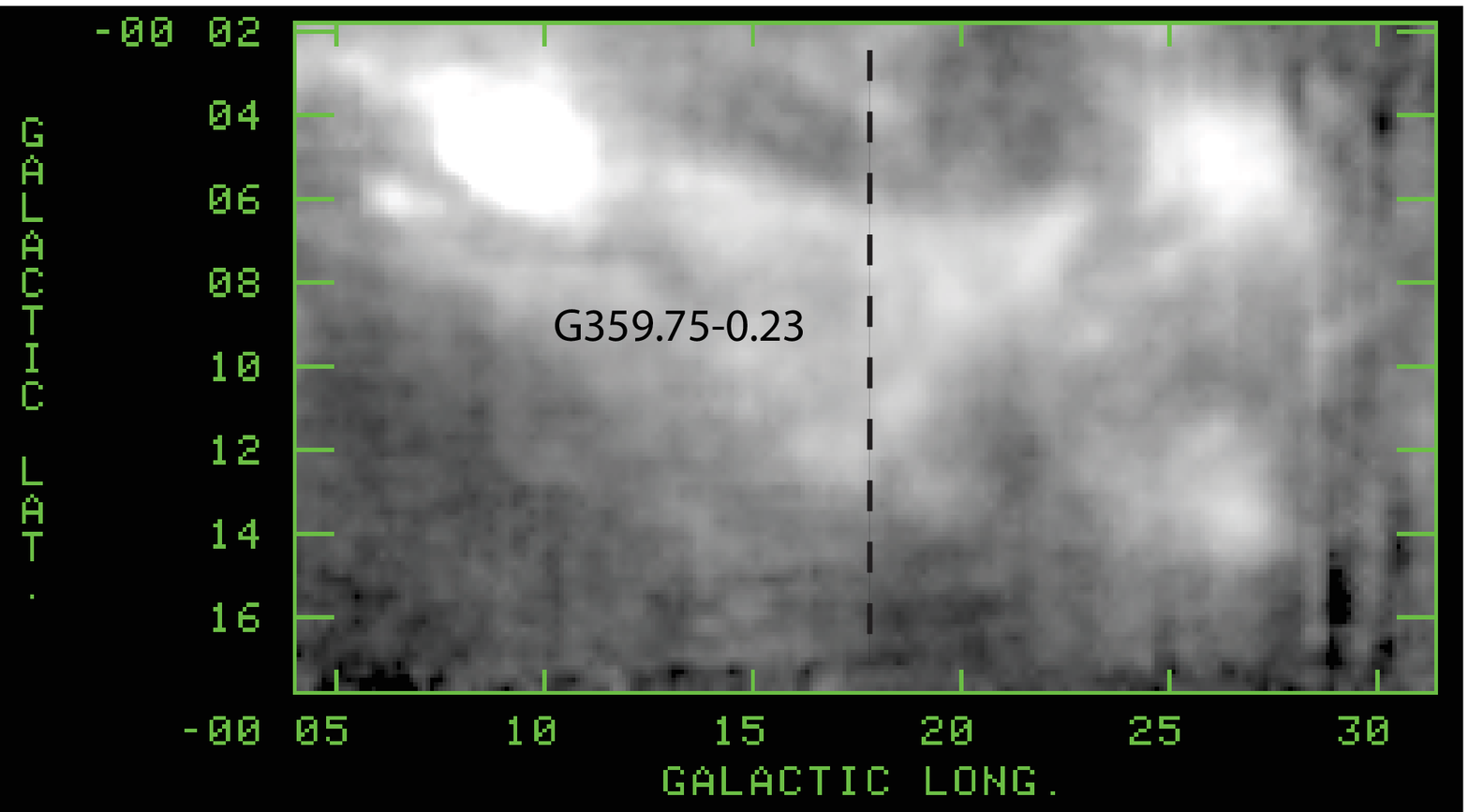}
\includegraphics[scale=0.45,angle=0]{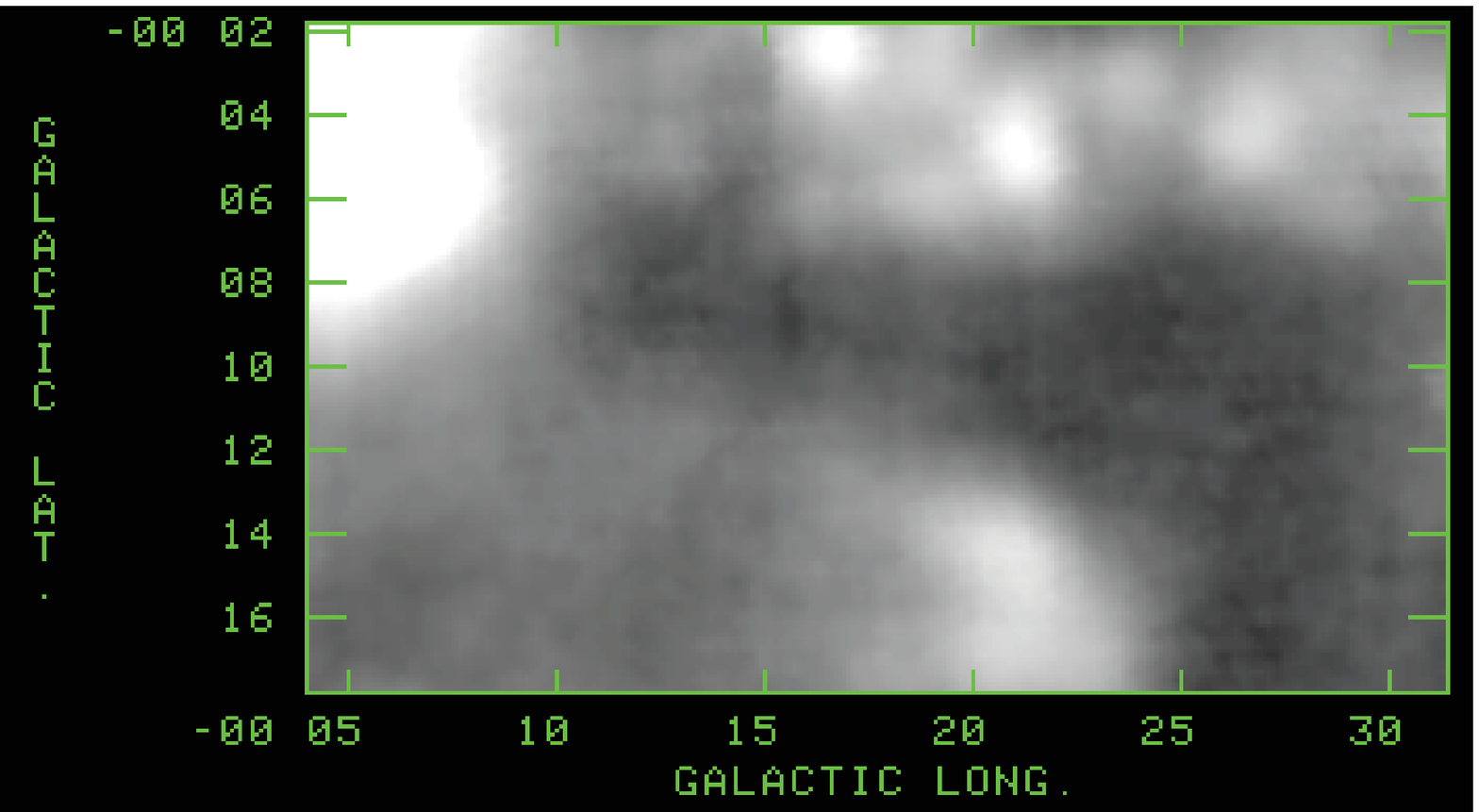}
\includegraphics[scale=0.45,angle=0]{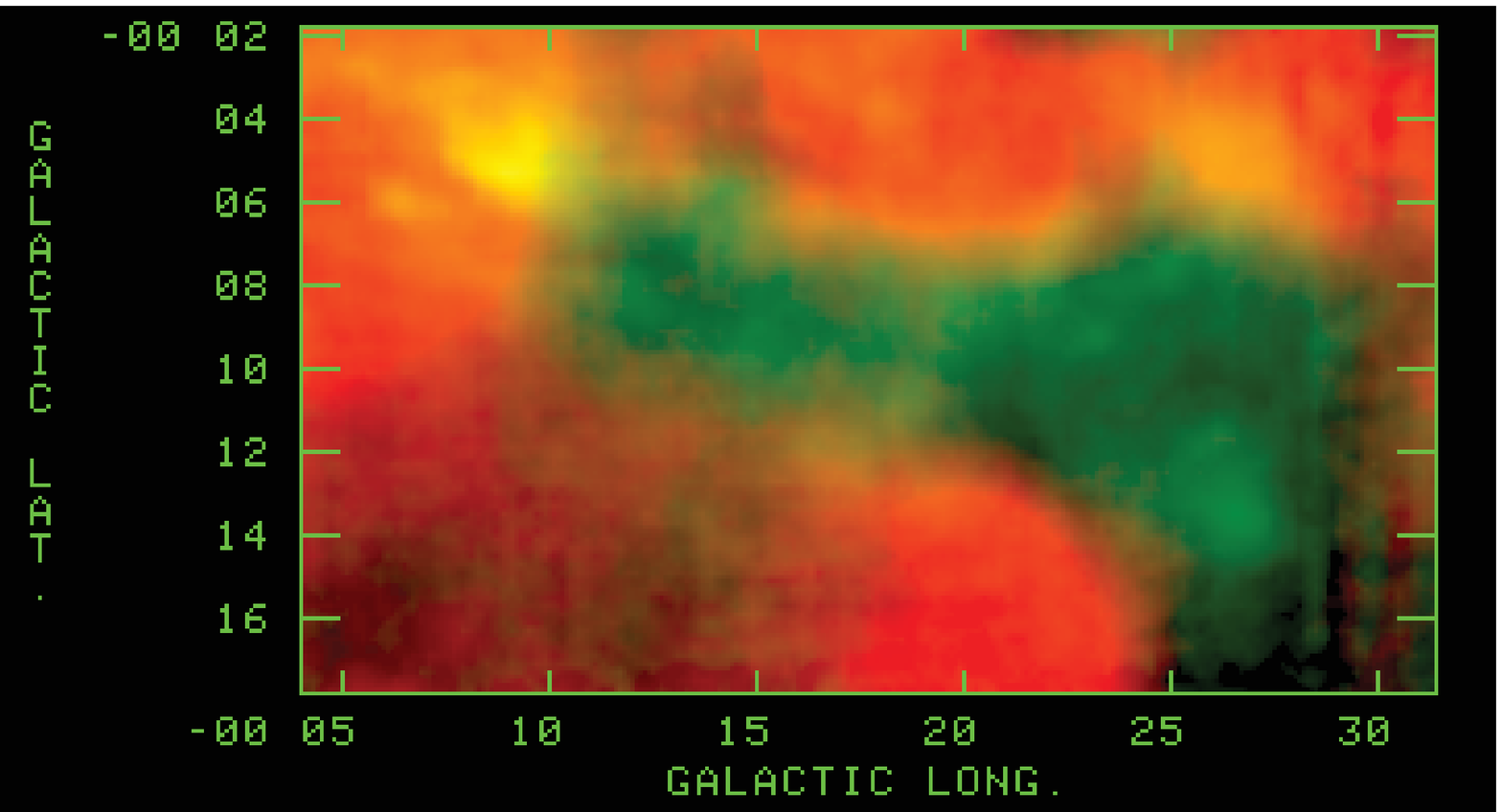}
\includegraphics[scale=0.25,angle=0]{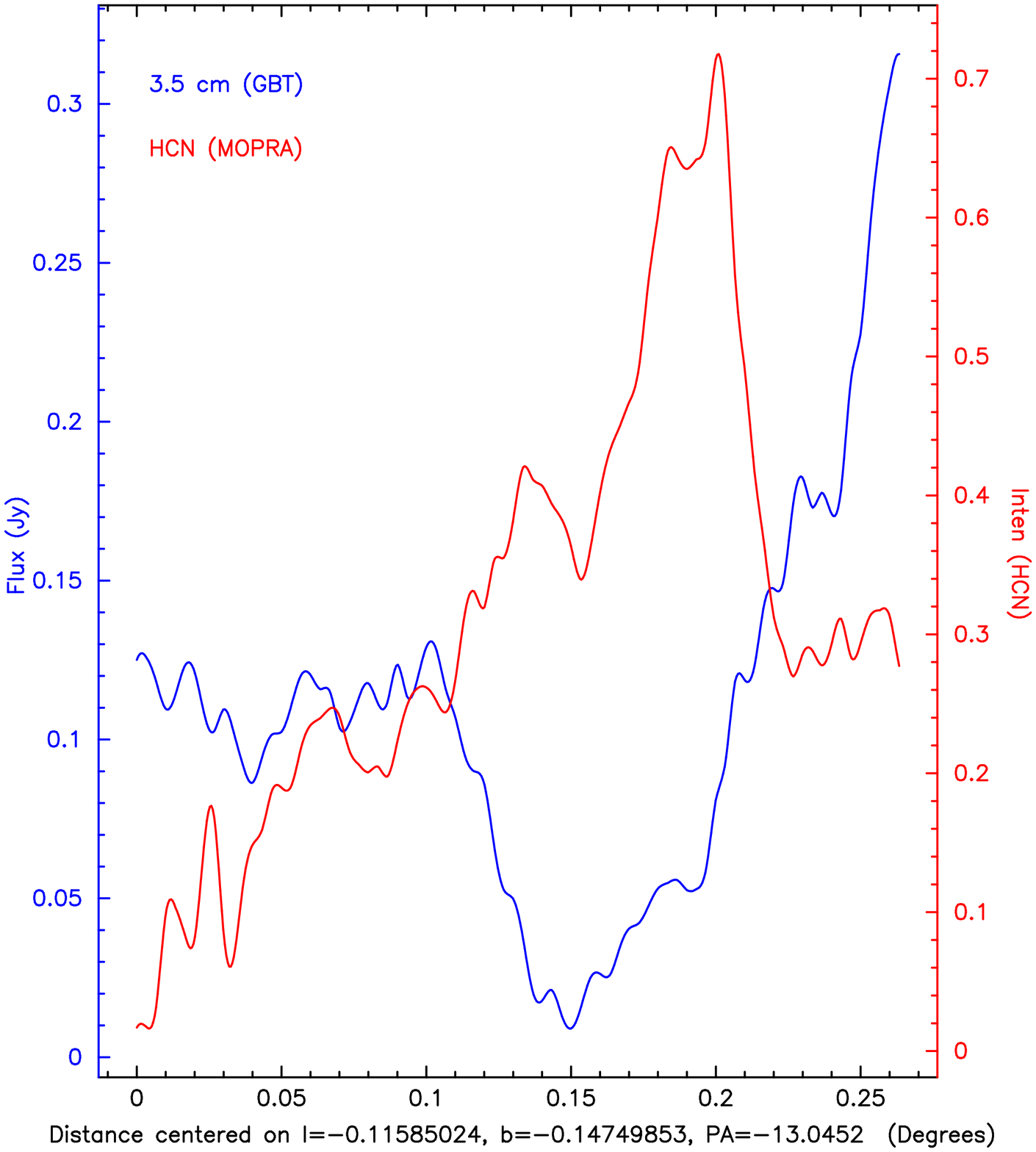}
\caption{
{\it (a) Top -}  
The integrated greysclae emission of the HCN (1-0) line  from 
 IRDC G359.75-0.13
over velocities  between -22.9 and -8.5  \kms\,  
based on MOPRA observations with a resolution of $\sim39''$.
{\it (b) Middle -}  
Similar to (a) except showing the distribution of 
radio continuum emission at 3.5 cm based on GBT observation with 
a spatial resolution of 88$''$.  
{\it (c) Bottom - } 
A color composite image showing (a) and (b) 
in red and green, respectively. 
{\it (d) Right - } 
A Cross cut, as drawn on (a),   made at constant l=16$'\,\,  36''$ 
showing the intensity profiles of HCN  line emission in red and 
and of 3.5 cm continuum data in blue. 
}
\end{figure}


\begin{figure}
\center
\includegraphics[scale=0.6,angle=0]{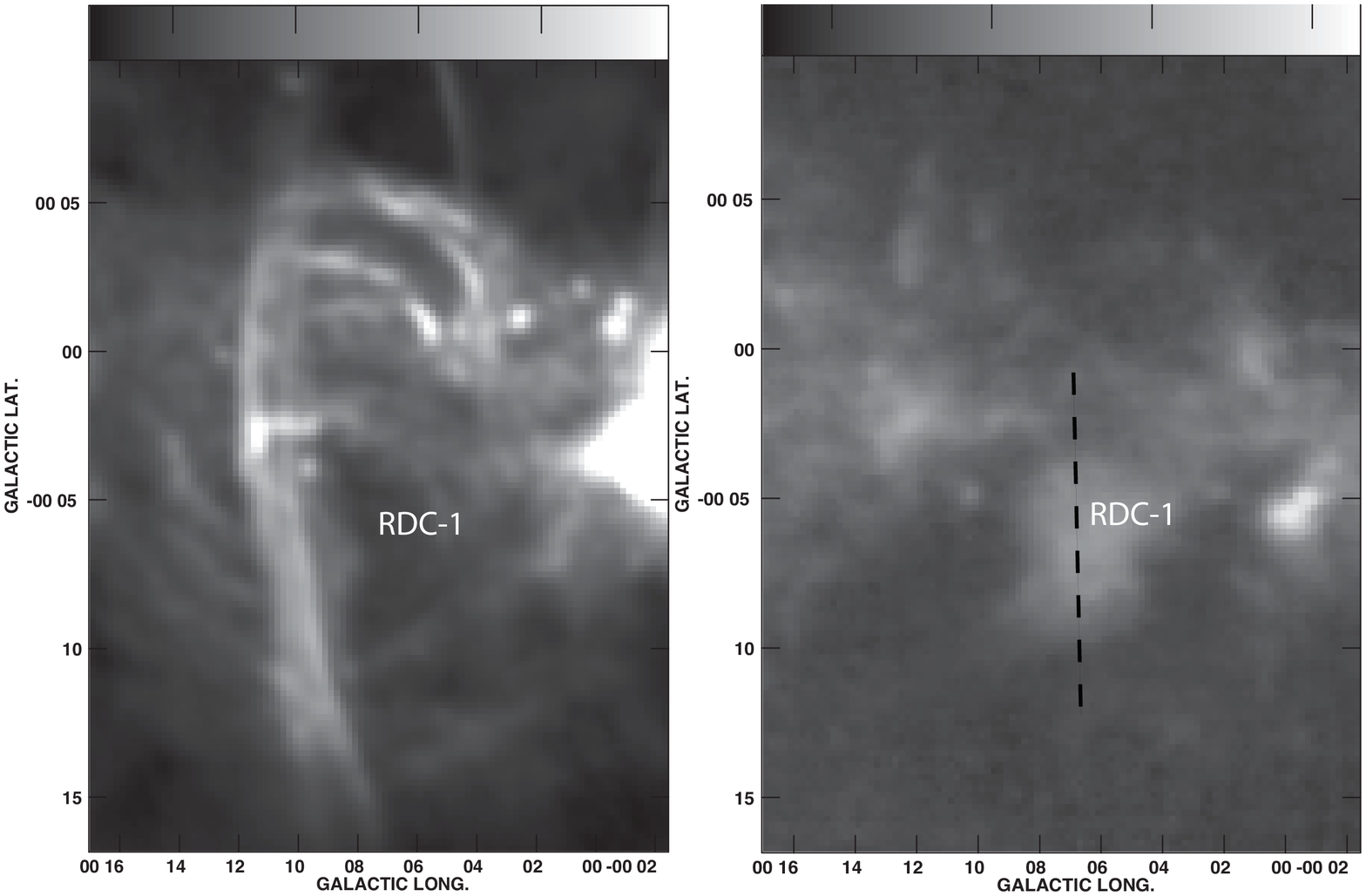}
\includegraphics[scale=0.6,angle=0]{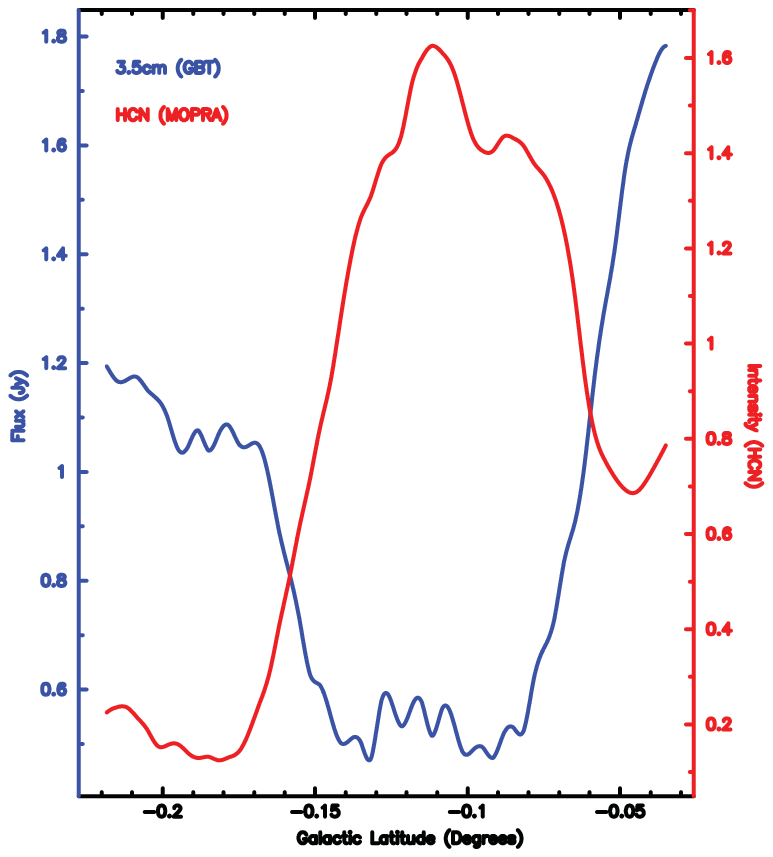}
\caption{
{\it (a) Top Left - } 
A  grayscale continuum image of the radio Arc  at 20cm with a resolution of 30$''$.
{\it (b) Top Right - } 
The integrated emission of the  HCN  line  over velocities between 0 and 50 \kms 
with a spatial resolution of 39$''$ (Jones et al. 2011). 
{\it (c) Bottom  Left - } 
Cross cuts aslong a line  drawn on (b)  are made at constant  l$\sim6'.1$
showing the profiles of HCN  line emission in red 
and of 3.5 cm continuum emission based on  GBT observation in blue.
}
\end{figure}

\begin{figure}
\center
\includegraphics[scale=0.3,angle=0]{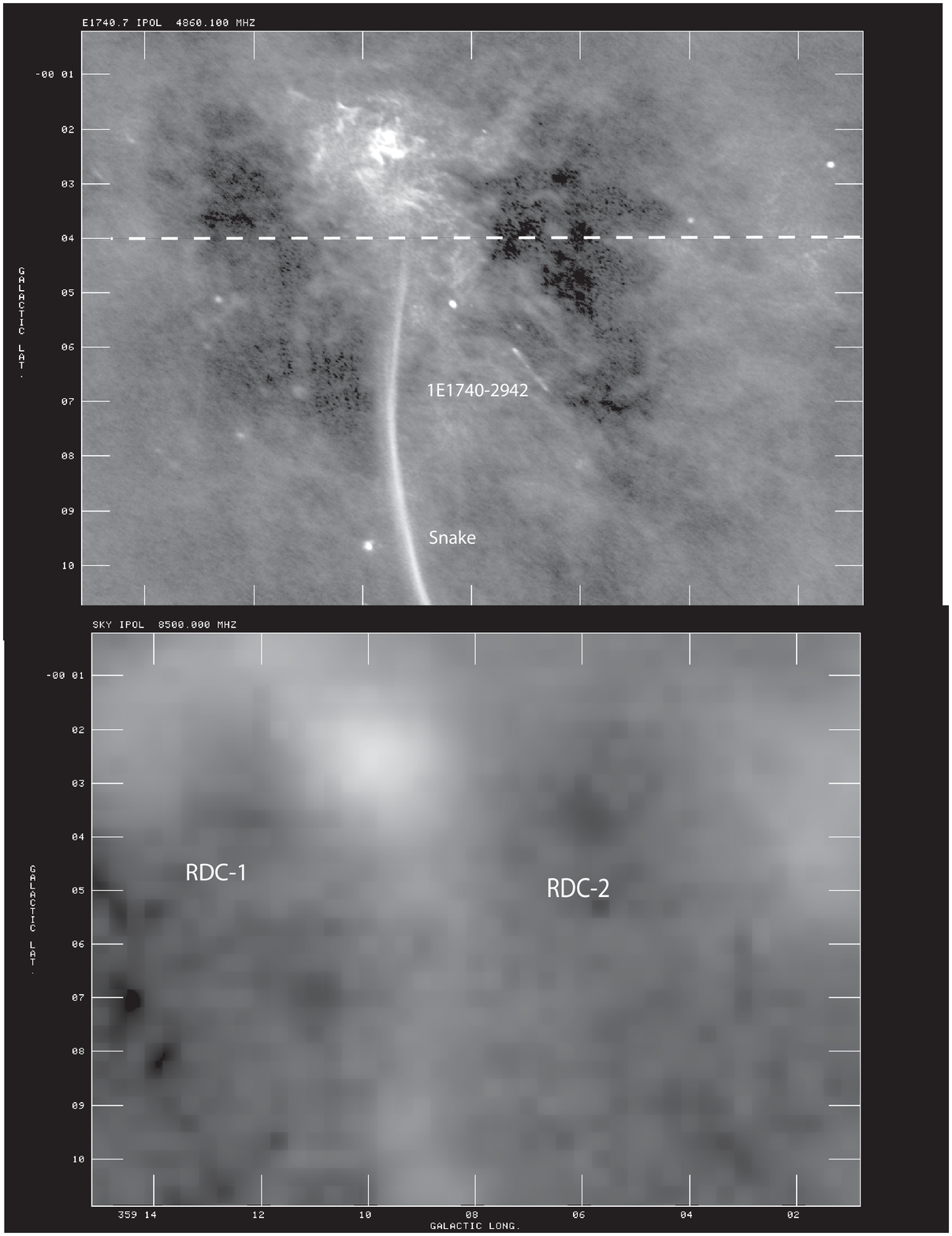}
\includegraphics[scale=0.4,angle=0]{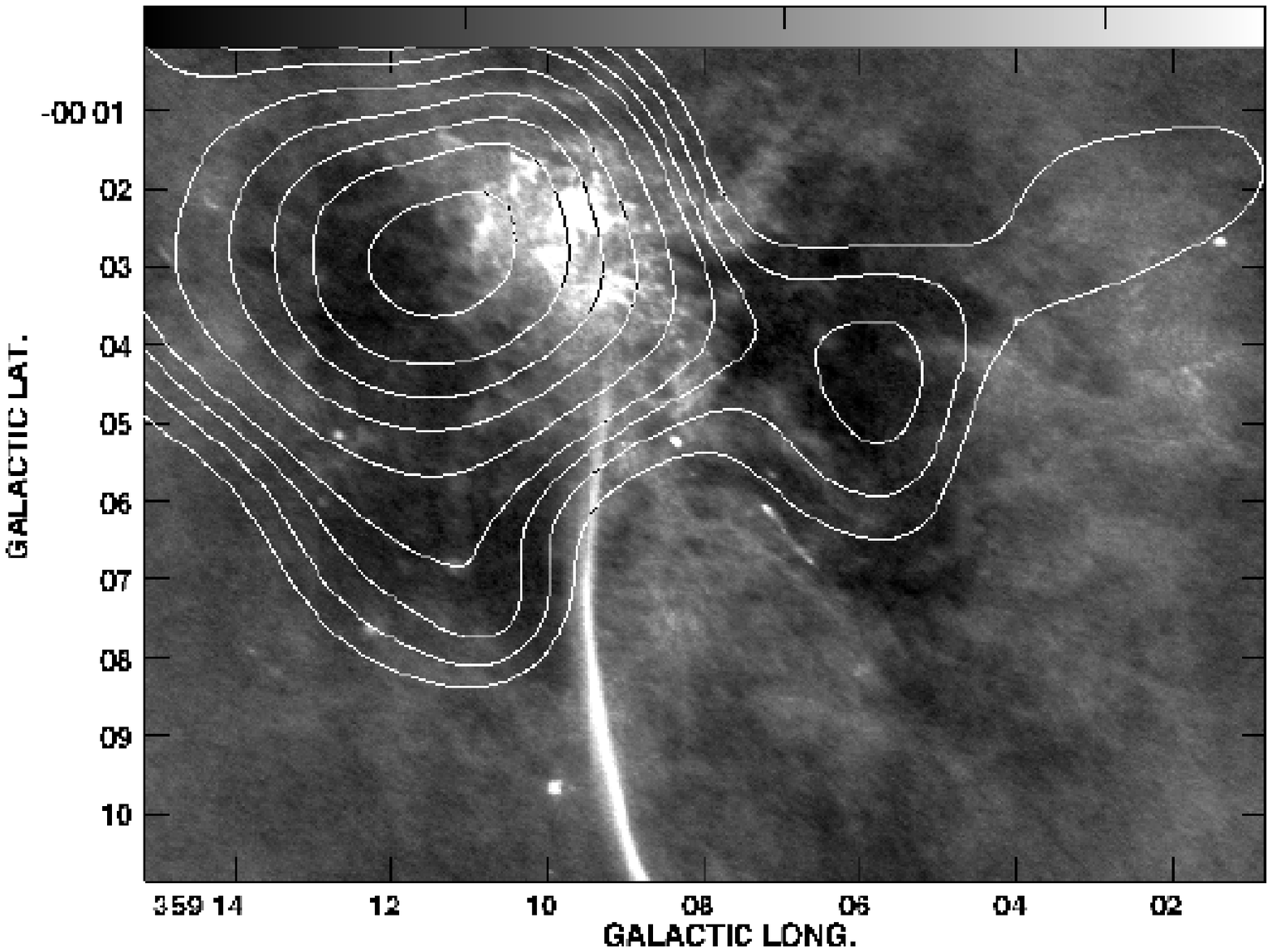}
\includegraphics[scale=0.3,angle=0]{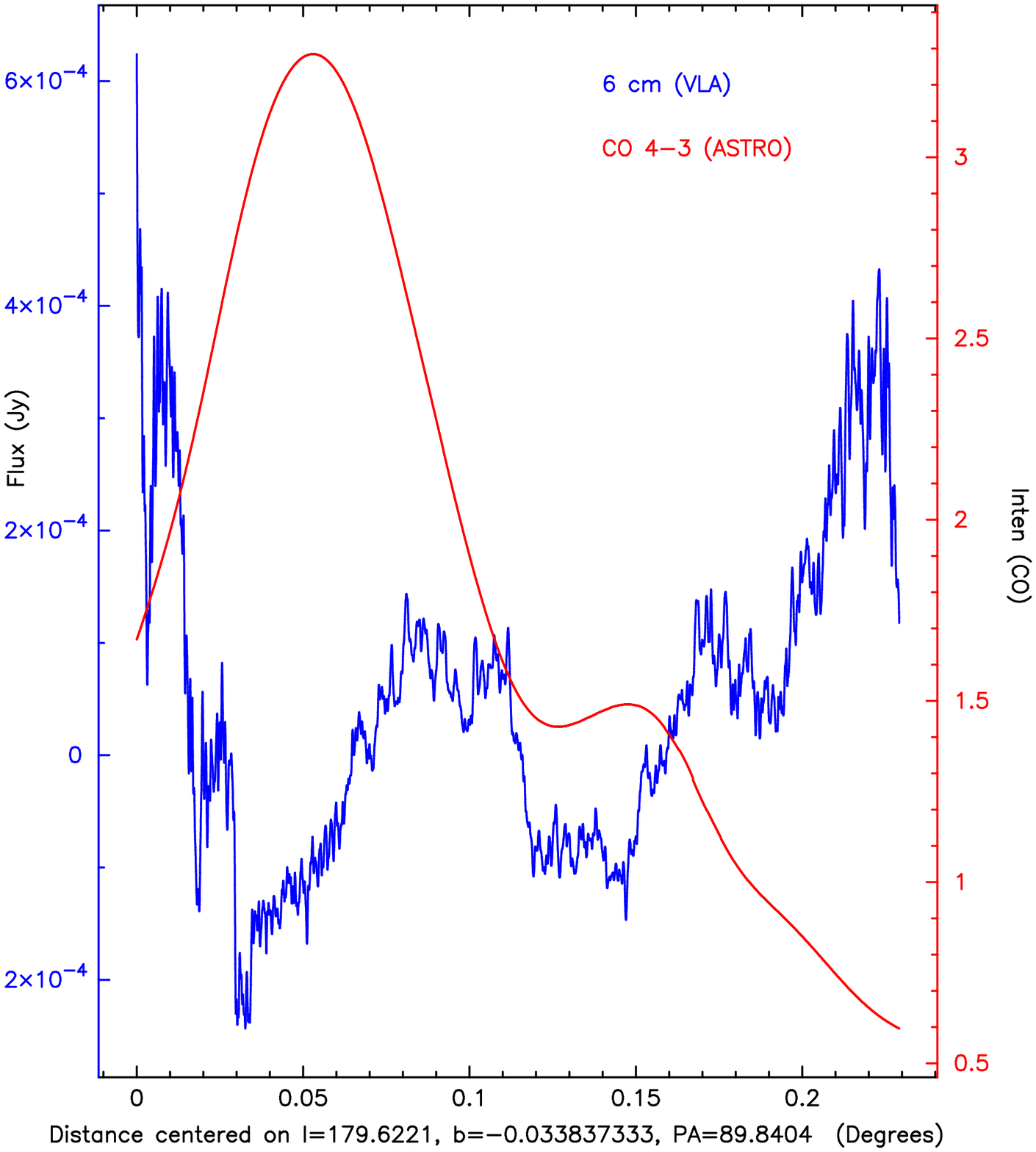}
\caption{
{\it (a) Top Left - }
A grayscale 6cm continuum image 
of the snake filament 
based on observations taken 
with all four configurations of the VLA with a 
resolution of  $3.72''\times1.86''$ and rms noise 10$\mu$Jy.
{\it (b) Bottom Left - }
Similar to (a) except that a 3.5 cm continuum image taken with 
the GBT  with a spatial  resolution of 89$''$.  
{\it (c) Top Right - }
A weighted average contours of CO (4-3) line emission integrated 
between  -150.5 and -140.5 \kms with a resolution of $\sim2'$ 
 superimposed on  (a). 
{\it (d) Bottom - }
Cross cuts along a line drawn on (a)
at b=$-3'\, 45''$  
show two CO  peaks and  two 6cm continuum dips, corresponding 
to   RDC-1 and RDC-2. 
The VLA image at 6cm   is primary beam corrected. 
}
\end{figure}


\begin{figure}
\center
\includegraphics[scale=0.5,angle=0]{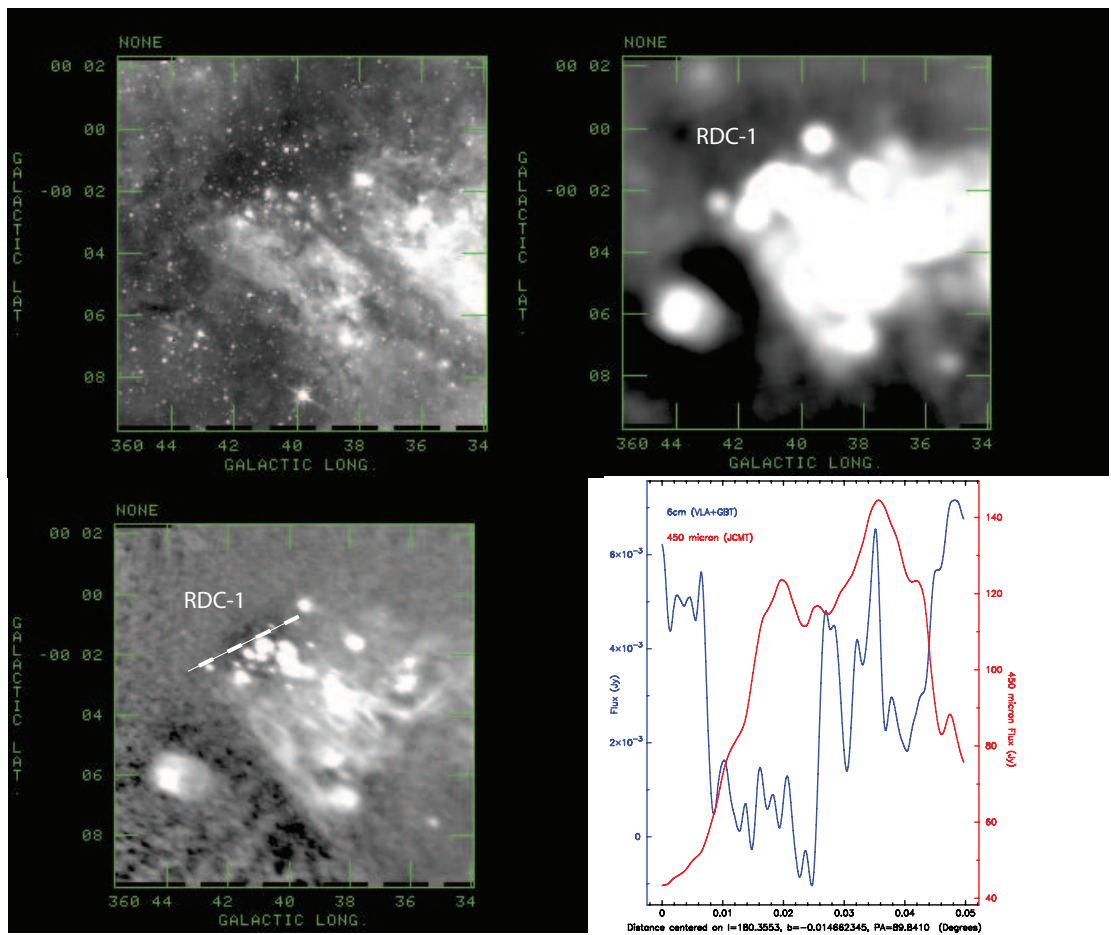}
\caption{
{\it (a) Top Left - } An IRAC image of Sgr B2 at 8$\mu$m 
with a  resolution of $\sim2''$
{\it (b) Top Right - } 
A continuum image of Sgr B2
at 20cm based on 
combining data taken from the C and D configurations of the 
VLA and the GBT  before the image was convolved to a 
resolution of 30$''$.  
{\it (c) Bottom Left - } Similar to (b) 
(both VLA and GBT data  combined) but at 
a resolution of  10.8$''\times5.5''$
{\it (d) Bottom Right - }  Cross cuts
centered  at $l=41'\, 15''\, b=-1'\, 12''$
along a line drawn on (c) show profiles of emission 
at 450$\mu$m and 6cm in  blue and red, respectively.
}
\end{figure}


\begin{figure}
\includegraphics[scale=0.35,angle=0]{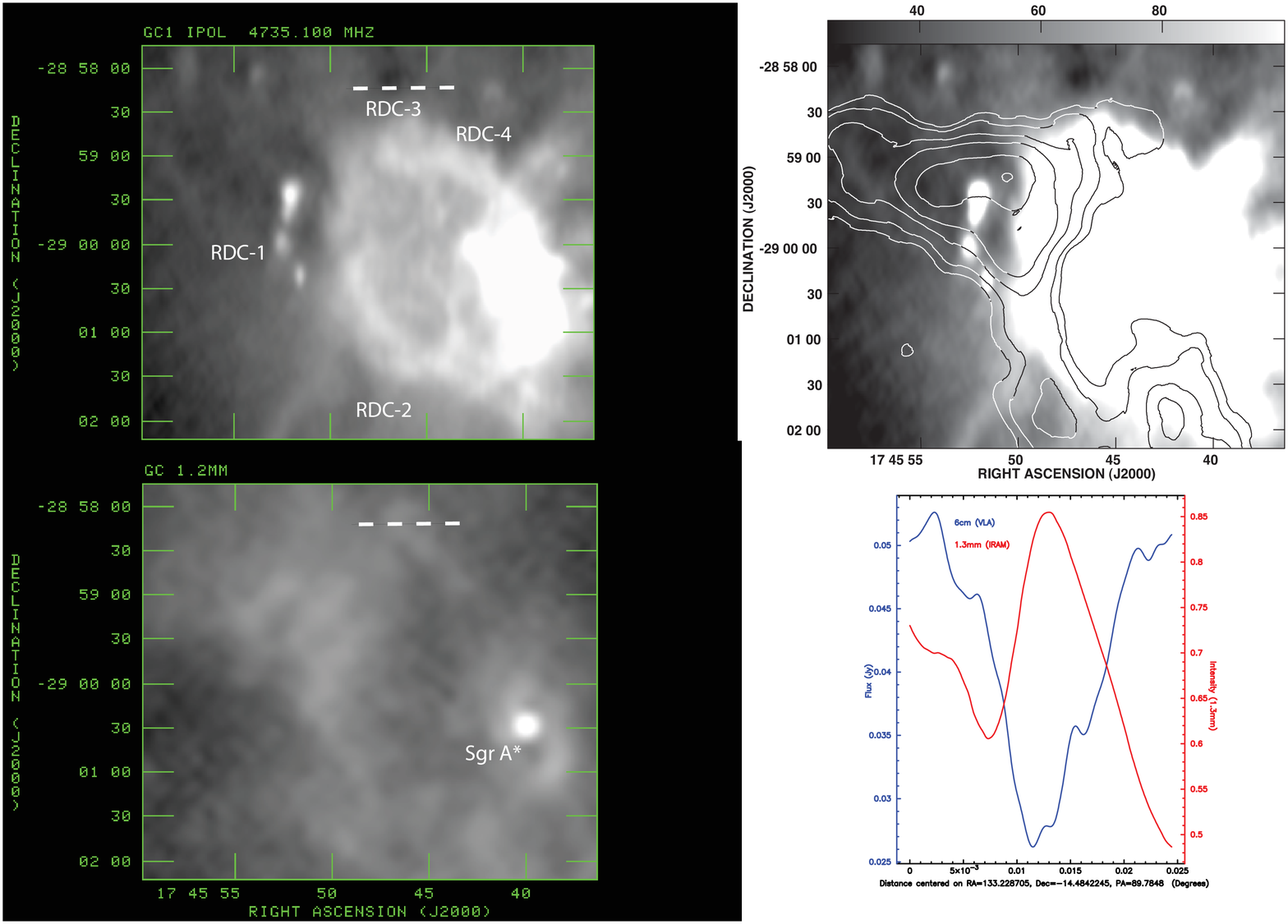}
\includegraphics[scale=0.35,angle=0]{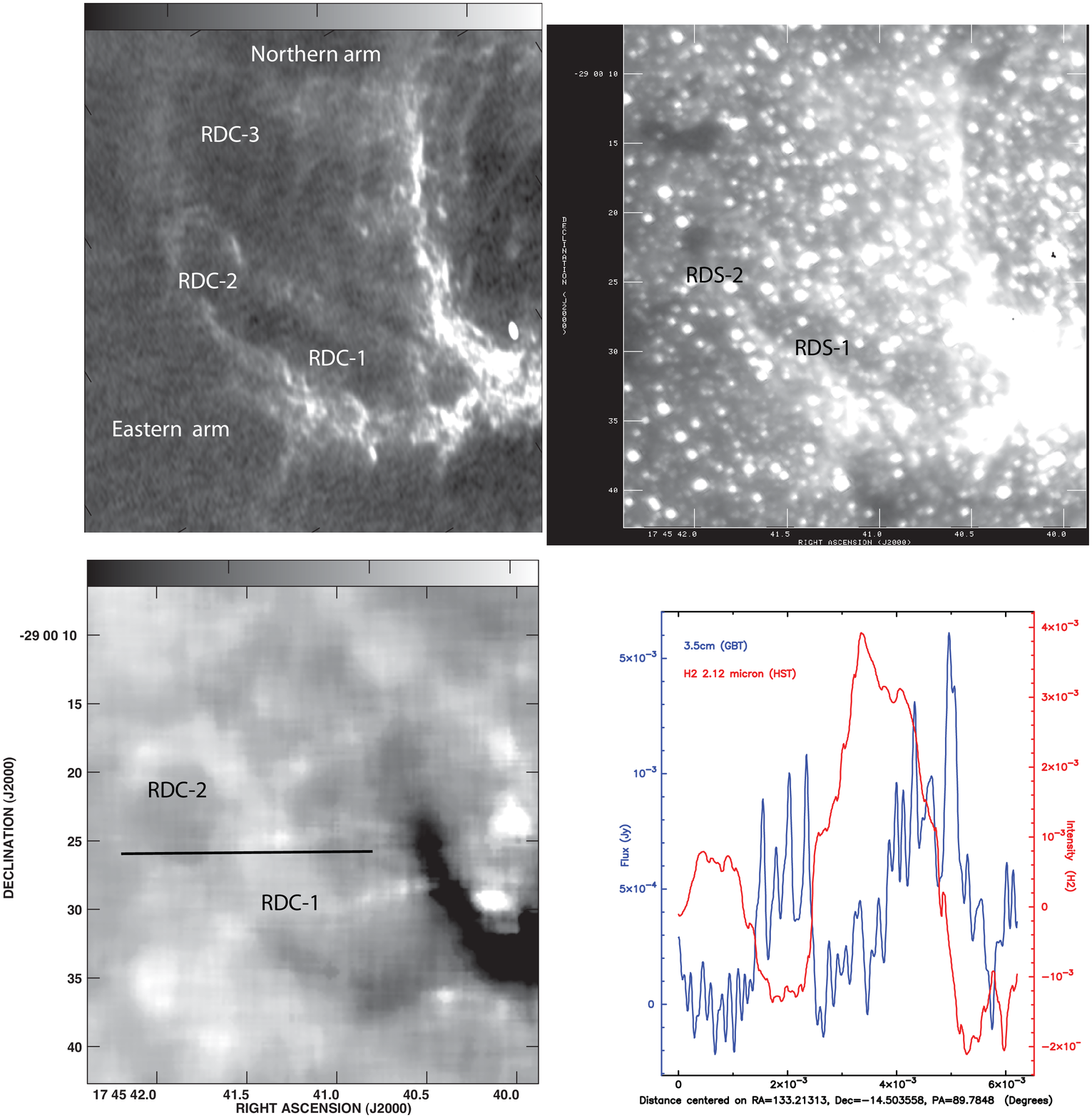}
\caption{
{\it (a-d) Top 4 - } Radio continuum of Sgr A East at 6cm,  the distribution of 
dust emission at 1.3 mm, contours of SiO (2-1)  line emission superimposed on  a 6cm continuum image 
and the cross cuts at 1.3mm and 6cm at the position of RDC-3 along the line drawn on (a) and (b).
{\it (e-h) Bottom 4 - } Radio continuum at 3.5cm wth a resolution of 0.45$''\times0.22''$, 
stellar subtracted H2 1-0 S(1) emission at 2.12$\mu$m, 
 an image of 1.87$\mu$m emission from stars and ionized gas, 
 the cross cuts   along the line drawn on (f)
show the emission profiles of  3.5cm and molecular H2 line.  
}
\end{figure}


\begin{thebibliography}{99}
\bibitem[]{}

\bibitem[]{}
Andersen, L. D., Snowden, S. L. \& Bania, T. M. 2010, ApJ, 721, 1319

\noindent
\bibitem[]{} 
Arendt, R.~G.,  {Stolovy}, S.~R., {Ramirez}, S.~V. et al. 2008, ApJ,  804, 



\bibitem[]{}
De Pree, C.G., Wilner, D.J., Goss, W.M. et al.  2000, ApJ, 540, 308

\bibitem[]{}
Egan, M. P., Shipman, R. F.,  Price, S. D. et al. 1998, ApJ, 494, 199

\bibitem[ferriere]{ferriere}
Ferriere, K.  2012, A\&A, 540, 50

\bibitem[]{}
Gray, A.D., Nicholls, J., Ekers, R.D. \& Cram, L.E. 1995, ApJ, 448, 164

\noindent
\bibitem[]{} 
Jackson, J. M., Geis, N.,  Genzel, R. et al.  1993,  ApJ, 402, 173	

\noindent
\bibitem[]{} 
Law, C. J., Yusef-Zadeh, F., Cotton, W. D., \& Maddalena, R. J. 2008, ApJS, 177, 255

\noindent
\bibitem[]{}
Lang, C. C., Goss, W. M., Cyganowksi, C., \& Clubb, K. I. 2010, ApJS, 191, 275L

\bibitem[]{} Jones, P. A., Burton, M. G., Tothill, N. F. H. \& Cunningham, M. R. 2011, MNRAS, 411, 2293


\bibitem[]{}
Lis, D. C. \&  Carlstrom, J. E. 1994, ApJ, 424, 189	

\bibitem[]{}
Martin, C. L.,  Walsh, W., Xiao, K. et al. 2004, ApJS, 153, 395

\bibitem[]{}
Mirabel, I. F., Rodriguez, L. F., Cordier, B. et al. 1992, ApJ, 358, 215

\noindent
\bibitem[]{}  
Molinari, S., Bally, J., Noriega-Crespo, A., \etal 2011, ApJ, 735, 33

\noindent
\bibitem[]{}  Morris, M. \& Serabyn, E. 1996, ARA\&A, 34, 645

\noindent
\bibitem[]{}  Nord, M. E., Lazio, T. J. W., Kassim, N. E., Hyman, S. D., LaRosa, T. N., et al. 2004, AJ, 128, 1646


\noindent
\bibitem[]{} Oka, T., Geballe, T. R., Goto, M., Usuda, T., \& McCall, B. J. 2005, ApJ, 632 882

\noindent
\bibitem[]{}  Pierce-Price, D., Richer, J. S., Greaves, J. S.,  \etal 2000, ApJ, 545, 121

\bibitem[]{} Reich, W. 2003, A\&A, 401, 1023

\bibitem[]{} Reid, M. Menten, K. M., Zheng, X. W.  et al. 2009, ApJ, 705, 1548

\noindent
\bibitem[]{} 
Sawada, T., Hasegawa, T., Handa, T. \& Cohen, R. J.  2004, MNRAS, 349, 1167

 
\noindent
\bibitem[]{} 
Tsuboi, M.,  Tadaki, K.,  Miyazaki, A. \&  Handa, T. 2011, PASJ, 63, 763	

\noindent
\bibitem[]{} 
Tsuboi, M., M., Ukita, N., \& Handa, T. 1997, ApJ, 481, 263


\noindent
\bibitem[]{} Uchida, K. I., Morris, M., Serabyn, E., G\"usten, R. 1996, ApJ, 462,  768 



\noindent
\bibitem[]{}  
Yusef-Zadeh, F., Hewitt, J. W., \& Cotton, W. 2004, ApJS, 155, 421

\noindent
\bibitem[]{}  
Yusef-Zadeh, F.,  Hewitt, J. W.,  Arendt, R. G. et al. 	 2009, ApJ, 702, 178	

\noindent
\bibitem[]{}  
Yusef-Zadeh, F., Hewitt, J. W., Wardle, M. et al. . 2012, ApJ, submitted


\noindent
\bibitem[]{}  
Yusef-Zadeh, F.  and Morris, M.  1987, ApJ,  322, 721

\noindent
\bibitem[]{}  
Yusef-Zadeh, F.,  Stolovy, S. R.,  Burton, M. et al. 2001, ApJ, 560, 749	

\noindent
\bibitem[]{}  
Zylka, R.,  Mezger, P. G. \&  Lesch, H. 1992, A\&A, 261, 119

\end{thebibliography}
\end{document}